\documentclass[aps,prl,twocolumn]{revtex4}
\usepackage{times}
\usepackage{graphicx}

\begin{document}
\title{Reconstructing the Density of States by History-Dependent Metadynamics}
\author{Cristian Micheletti$^1$, Alessandro Laio$^2$ and Michele Parrinello$^2$}
\affiliation{
(1) International School for Advanced Studies (SISSA) and INFM,
Via Beirut 2-4, 34014 Trieste, Italy\\
(2) Physical Chemistry ETH, H\"onggerberg HCI, CH-8093 Zurich, Switzerland.}
\date{\today}

\begin{abstract}
We present a novel method for the calculation of the energy density of states  
$D(E)$ for systems described by classical statistical
mechanics. The method builds on an extension of a recently proposed
strategy that allows the free energy profile of a canonical system to
be recovered within a pre-assigned accuracy,[A. Laio and
M. Parrinello, PNAS 2002].  The method allows a good control over the error on the recovered system entropy.
This fact is exploited to obtain $D(E)$ more efficiently by
combining measurements at different temperatures. The accuracy and
efficiency of the method are tested for the two-dimensional Ising
model (up to size 50x50) by comparison with both exact results and
previous studies. This method is a general one and should be
applicable to more realistic model systems.
\end{abstract}
\maketitle

It has long been recognized that all energy-related thermodynamic
properties of a classical canonical system can be calculated once the
energy density of states $D(E)$ is known. In fact, starting from the
partition function $ Z=\int dE\,D(E)e^{-\beta E}$, quantities like
free energies, specific heats and phase transition temperatures can be
computed in a straightforward manner. In principle, all canonical
averages can be calculated through a multidimensional density of
states. Due to this central role in equilibrium thermodynamics 
a variety of theoretical and computations studies have addressed the
problem of how to obtain $D(E)$ (or, equivalently, the entropy profile
$S(E) = \ln (D(E)$) in a reliable and efficient way.
 
In principle, $D(E)$ could be calculated from the histogram of the
energies visited during a single ``very long'' constant-temperature
simulation\cite{bennett}. In practice, for any finite simulation only
a limited energy windows is sampled so that the recovery of the system thermodynamics over a wide temperature
range is unfeasible.

Several alternative strategies have been developed to remedy this
shortcoming. For example the multiple histogram reweighting technique
relies on performing several simulations at different temperatures, so
as to explore different (overlapping) energy intervals
\cite{multihisto}. The
various histograms are then optimally combined to obtain $D(E)$ over
the union of the energy intervals. 
Another successful family of techniques aims at obtaining $D(E)$ by
changing iteratively the probability with which the various energy
levels are visited in stochastic dynamics until the recorded energy
histogram is ``flat''. Such methods include entropic sampling
\cite{Lee}, multicanonical and broad-histogram techniques techniques
\cite{muca,bhm} and also the recent method of Wang and Landau
\cite{wl01}.

These and similar techniques have proved to be very valuable in a
variety of contexts \cite{hansmann,hesse,TROW}, but there is still
ample scope for alternative approaches that could provide improved
efficiency and better error control. Here we propose to modify and
extend the metadynamics method recently introduced by two of us
\cite{hills} for evaluating $D(E)$.  The algorithm we introduce allows
a good error control on the explored energy range.  Moreover, although
within our approach $D(E)$ could be reconstructed performing
simulations at virtually any temperature, it is still possible to
exploit the Boltzmann bias to focalize the computational effort for
exploring the region of phase space of relevance for a temperature of
interest (e.g. of a phase transition).
In the following we will first describe this general strategy
followed by its application to the typical reference case constituted
by the two-dimensional Ising model.

The essence  of the metadynamics approach is first to identify those
relevant collective variables (CVs) that are difficult to sample.  In
other applications\cite{hills} a similar metodology was applied in
order to observe a specific transition (e.g. a chemical reaction) in
systems described by a complex atomistic Hamiltonian.  At this scope
CVs explicitly depending on the microscopic configuration of the
system have been employed, like, e.g., coordination numbers or
distances between specific atoms of the system.  Here, we aim at
reconstructing the canonical free-energy profile, $F(E) =E-T\,S(E)$
and the relevant variable is $E$ (which is also an explicit function
of the microscopic configuration of the system).  At each metadynamics
step the system evolution is guided by the generalised force which
combines the action of the thermodynamic force (which would trap the
system in free-energy minima) $dF(E) / dE$ and a history-dependent
one which disfavours system configurations already visited.  The
history-dependent potential, $F_G$, is constructed as a sum of
Gaussians of width $\Delta E$ and height $w$ centred around each 
value of $E$ already explored during the
dynamics.  As shown in ref \cite{hills}, $F_G$ fills in time the
minima in the free energy surface and, in the limit of a long
metadynamics $F(E)+F_G(E)$ tends to become flat as a function
of $E$ and hence $-F_G(E)$ becomes an approximant of $F(E)$.

Clearly, the exact form of the metadynamics equations
and the choice of the parameters $w$ and
$\Delta E$ may affect significantly the accuracy of this
estimate. Furthermore careful control of the error is essential for
reconstructing a reliable density of states. This requires that the
algorithm in ref. \cite{hills} be substantially improved.
The modified metadynamics equations are:

\begin{eqnarray}
E_G^{t+1} = E^{t} + \Delta E \frac{f\left( E^{t}\right) }{\left| f\left(E^{t}\right) \right|}
\label{aa1}   \\
E^{t+1}=E^{t}+\Delta E (1+\frac \chi 2)\frac{f\left( E^{t}\right) }{\left| f\left(
E^{t}\right) \right| } \label{a1} 
\end{eqnarray}
where $\chi$ is a random number between 0 and 1. 
The generalized force $f(E)$ is given by $ f\left(
E\right) =-{\partial \over \partial E} \left[ F( E) +F_{G}^{t}(E)
\right] $ with the history-dependent potential, $F^t_G$, defined as
\begin{equation}
F^t_{G}\left( E\right) =\sum_{u\leq t}w\,e^{-\frac{(
E-E^{u}_G)^{2}}{2\Delta E^{2}}} \ .
\label{a3}
\end{equation}
By displacing  the center of the Gaussian $E_G^{t+1}$ with respect to 
the point of evaluation of the
generalised force (equation \ref{aa1}), the added Gaussian maximally
compensates $f(E^t)$ and flattens $F+F_G$ around $E^t$.
The energy step performed at every metadynamics iteration is chosen 
randomly in the interval between $\Delta E$ and $ 1.5 \Delta E$ (equation \ref{a1})
in order to reduce the correlation induced by the dynamics.

\begin{figure}[h]
\includegraphics[width=3.0in,height=3.0in]{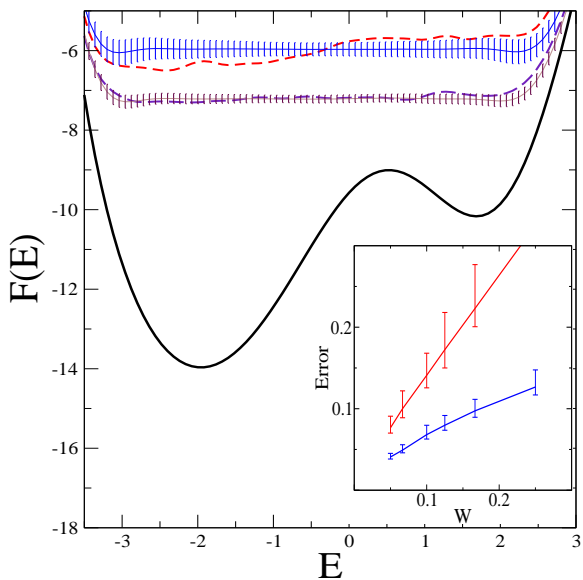}
\caption{Example of reconstruction of a preassigned (analytic)
free-energy profile (black curve), $F(E)$, by means of metadynamics runs
consisting of 200 Gaussians of spread $\Delta E=0.4$ and height
$w=0.16$. To mimic the uncertainty on the thermodynamic force
encountered in stochastic simulations, a Gaussian noise of width 0.3
was added to $F^\prime(E)$.  The broken red and indigo lines denote
the filled profile, $\delta F(E) = F(E) - F_R(E)$, obtained
respectively with an unsmoothed ($\tau_c =0$) and smoothed ($\tau_c
=100$) metadynamics. Notice the shorter correlation lengths and
smaller amplitudes of the fluctuations in the second case.  The
average and dispersion of $\delta F(E)$ were calculated (blue and
brown lines wth error bars) for 1000 independent runs, revealing both
the absence of biases in $F_R(E)$ and the constancy of the error
except close to the boundaries. By changing $w$ and modifying the
number of Gaussians so as to work at a constant filled free-energy
volume, we calculated over the range $|E|<2.5$ the average, maximum
and minimum values of $\delta F(E)$ measured over 1000 runs, see
inset. The linear dependence of the error on $F_R$ is thus apparent
both for the unsmoothed and smoothed metadynamics (red and blue
respectively).}
\label{fig:fig1}
\end{figure}

Finally, in order to further reduce the spatial correlations in $F_{G}$, we notice that
when the metadynamics is terminated, say at position $E^{t}$, $F_{G}$ will
present a bump in a region around $E^{t}$ whose spread depends on the
correlation time of the metadynamics.  This effect can be lessened if
the contribution of the Gaussians placed at the end of the dynamics
are weighted less. Therefore, after the metadynamics with constant $w$
is terminated we reconstruct the free energy from $
F_{R}\left(E\right) =-\sum_{u\leq t}w\tanh \left( \frac{ t-u}{\tau
_{c}}\right) e^{-\frac{\left( E-E^{u}\right)^{2}}{2\Delta E^{2}}}$
where $\tau _{c}$ is taken to be larger than the typical time required
to sweep the ``filled'' energy range.  Other smoothing functions, such
as $\min(1,\frac{t-u}{\tau _{c}})$ can, of course, also be chosen and
have an analogous effect.

The modified metadynamics algorithm allows the efficient
reconstruction of $F(E)$ in the explored energy range, within an error
that is ultimately controlled only by the Gaussian's height $w$.  We
demonstrate the quality of the algorithm by reconstructing an
``ideal'', pre-assigned, $F\left( E\right)$ (see
Fig. \ref{fig:fig1}). If the method is void of systematic biases one
would expect the quantity $\delta F\left( E\right) \equiv F_{R}\left(
E\right) -F\left( E\right)$ to be, on average, constant throughout the
filled energy range. Moreover, we would expect deviations from the
constant average value to be of the order of $w$. These properties are
confirmed by the results presented in the inset of Fig. 1, where the
uniformity of the average value of $\delta F(E)$ is apparent, together
with the constancy of its dispersion, $\sigma ^{2}\left( E\right)
=\langle\delta F\left( E\right) ^{2}\rangle -\langle \delta F\left(
E\right) \rangle ^{2}$. The plots also illustrate the benefits of the
``smoothing'' procedure over the last part of the metadynamics
trajectory, since this results in a decrease of the spread of $\delta
F(E)$. As is visible in the inset of Fig. 1, the dispersion is further
confirmed to be approximately proportional to $w$. An important fact is
that near the boundaries of the explored energy interval $F_R(E)$
decays to zero and hence deviates from the true free energy. To
identify the interval over which $F_R$ is reliable we need to
ascertain if the number of Gaussians accumulated at a given energy
$E$, $\approx |F_R(E)/w|$, is significantly larger than the minimal
number of superposed Gaussians needed to produce the observed free
energy derivative, $\approx F^\prime(E)/(w/\Delta E)$. In this work we
have required that the ratio of the former to the latter be greater
than 5.

We now use the algorithm described above to compute $D(E)$ for an
$N$x$N$ two-dimensional Ising model with ferromagnetic
nearest-neighbour interactions and periodic boundary conditions
\cite{ising}. In fact, exact expressions for S(E) are
available\cite{beale} and the error induced by the algorithm can be
explicitly estimated and comparison with other approaches can be
made\cite{wl01}.

Virtually the entire computational effort of the metadynamics is spent
in estimating the thermodynamic forces $\partial F/ \partial E$ at
each energy value, $\bar{E}$, visited by the metadynamics. To respect
the discrete nature of the system's energy spectrum, the continuous
value of $E^t$ produced by eq. (\ref{a1}) was discretized to the
nearest energy level. Due to the discreteness of the energy
spectrum, the thermodynamic force in $\bar{E}$ cannot be estimated by
the Lagrange-multipliers technique described in Ref.\cite{hills}, and
is rather obtained using a centred difference approach. If $p_1$ and
$p_2$ are the occupation probabilities of the two energy levels, $E_1$
and $E_2$ adjacent to $\bar{E}$ (we assume $|\bar{E} - E_1| =|\bar{E}
- E_2|$), we have:

\begin{equation}
{\partial F \over \partial E}|_{E = \bar{E}} = {T \over (E_2 -
E_1)}\, \ln{p_1 \over p_2} \ .
\label{eqn:force}
\end{equation}

\noindent $p_1$ and $p_2$ are evaluated with an umbrella-sampling
strategy consisting of a Monte Carlo evolution of the system
(Metropolis acceptance/rejection of single-spin flips) under the
action of an effective Hamiltonian obtained by adding to the energy of
a given spin configuration, $E$, the term $1/2 \, K (E -
\bar{E})^2$. A suitable choice of the parameter $K$ forces the system
to explore the energy region around $\bar{E}$ thus populating
appreciably both levels $E_1$ and $E_2$. The symmetry with respect to
$\bar{E}$ of the added umbrella potential allows to calculate the
thermodynamic force through the same eq. \ref{eqn:force} with $p_1$
and $p_2$ being the fraction of times that the corresponding energy
levels are encountered in $n$ statistically independent configurations
picked with the modified canonical weight. For this purpose the Monte
Carlo trajectory was sampled at intervals comparable to the
autocorrelation time after having discarded a few tens of initial
system sweeps. 
The MC sampling was stopped when the
estimated uncertainty on the force \cite{note_binomial} was equal to
the maximum force introduced by a single Gaussian, $w \, \exp(-1/2) /
\Delta E$.  This choice ensures that, for large values of $t$,
$f(\bar{E})$ is of the order of $w/\Delta E$. 
If the force is calculated with much greater accuracy the repeated 
 superposition of the Gaussians would still lead to an uncertainty of order 
 $w/\Delta E$ on $f(\bar{E})$.

By means of such a metadynamics it is therefore possible to
reconstruct the free energy profile, $F_R(E,T)$. An estimator for the
system entropy is given by $S_R(E) = [E - F_R(E,T)]/ T$. The
uncertainty over $F_R(E,T)$ is inherited by $S_R(E)$ whose {\em a
priori} dispersion is thus of the order $w/T$.  Thus, the expected
error on the entropy profile is constant. This represents a major
difference over standard reweighting techniques, where the accuracy on
the calculated entropy usually deteriorates as one moves away from the
free energy minima.

If the goal is to reconstruct $S(E)$ over a wide range of energy, it
seems natural to combine the outcome of several meta-dynamics at
various temperatures, in analogy with multiple-histogram
techniques\cite{multihisto}.  In the following we shall indicate with
$S_{R,i}$ the entropy reconstructed in the $i$th metadynamics carried
out at temperature $T_i$ and with Gaussians of height $w_i$ and width
$\delta_i$. Due to the temperature-dependence of the free energy, each
run will typically explore a different energy range. The data obtained
in the different metadynamics runs can be optimally combined to
provide a single entropy estimate, $\tilde{S}$, over the union of the explored
regions\cite{multihisto}. To do so we recall that the entropy
$S_{R,i}$ is known only up to an additive constant $c_i$ and that the
uncertainty on $S_{R,i}$ is $\epsilon_i = w_i/T_i$ throughout the
reliably-explored energy range.

\begin{figure}[h]
\includegraphics[width=3.0in,clip=]{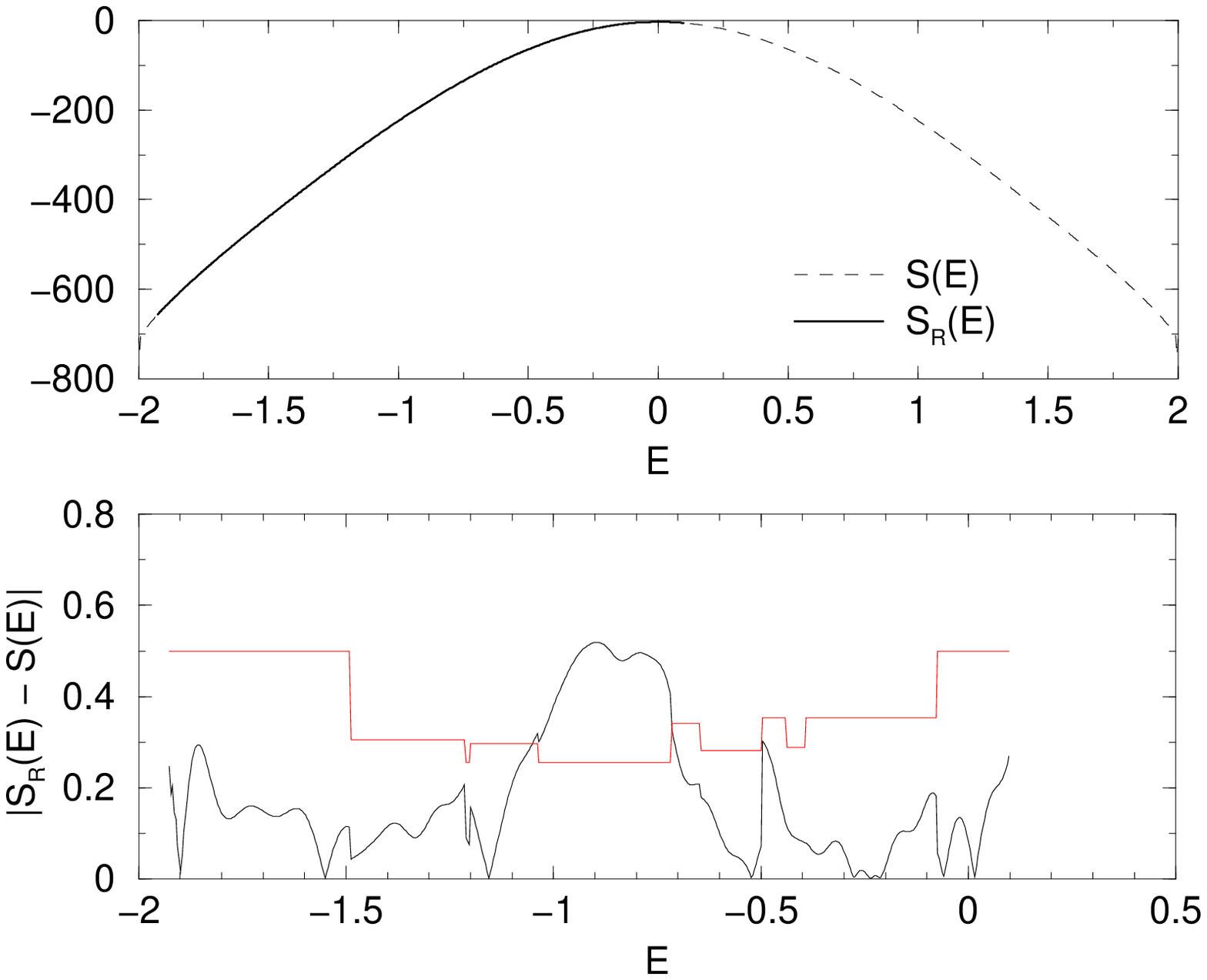}
\includegraphics[width=3.0in,clip=]{final_cv_32.eps}
\caption{Results from the metadynamics runs carried out on the 32x32
Ising system. Top: the exact and reconstructed entropies. In both
cases the entropy has been normalised so that $\sum_E \exp[S(E)] =1$.
The abscissa indicates the energy per spin. Middle: Difference between
the true and reconstructed entropy. The dispersion expected {\em a
priori} on the reconstructed entropy is shown with a red line. Bottom:
the exact and reconstructed specific heat and average energy (inset)
as a function of temperature.}
\label{fig:ising}
\end{figure}

\noindent This leads us naturally to consider a maximum likelihood
approach to obtain $\tilde{S}$ and the additive constants by
minimizing the least-squares function,

\begin{equation}
{\cal L}(\tilde{S},c_1,...c_n) = \sum_{i=1,n} \sum_E { | \tilde{S}(E)
 - S_{R,i}(E) -c_i|^2 \over \epsilon_i^2} 
\label{eqn:like}
\end{equation}

\noindent where the first sum is carried over the various metadynamics
runs and the second one runs over the system (discrete) energy levels
with the proviso that $\epsilon_i$ is equal to infinity outside the
reliable energy range. The determination of $\tilde{S}(E)$ through the
minimization of ${\cal L}$ relies on the statistical independence of
each term in the sum of equation (\ref{eqn:like}). This is realized
only approximately due to the existence of an intrinsic scale of
autocorrelation for the reconstructed free-energy/entropy dictated by
the correlation lenght of the metadynamics.  Therefore, the minimization of
(\ref{eqn:like}) is meaningful provided that each energy value is
covered by several metadynamics runs, each exploring an interval
substantially larger than the Gaussian widths.

The requirement of stationarity for ${\cal L}$ leads to
self-consistent equations which can be solved iteratively in terms of
the $c_i$'s and $\tilde{S}(E)$. Despite the presence of the additive
terms, $c_i$'s, which distinguish the present juxtaposition scheme
from others already available, the self-consistent equations are
simple both in their formulation and numerical
implementation. Convergence to the solutions is typically reached in a
few tens of iterations.  The
least-squares approach also allows the expected standard
deviation of $\tilde{S}(E)$ to be calculated:
\begin{equation}
\sigma(\tilde{S}(E))^{-1/2} = \sum_{i=1..n} \epsilon_i(E)^{-2}\ .
\label{eqn:error}
\end{equation}
The use of the reconstruction strategy is first illustrated
for the 32x32 Ising model, see Fig. \ref{fig:ising}.  The curve for
the entropy, Fig. \ref{fig:ising}, results from the combination of
runs at six temperatures, $T=$ 2, 2.6, 3.0, 3.4, 6.0 and 12.0. At each
temperature 1000 Gaussians where used with $w/T \approx 0.5$, $\Delta
E \approx 0.04$, $K\approx 0.4$ and $\tau_c=300$. The total
computational effort required $7.5\, 10^5$ MC sweeps.  The comparison
with the true system entropy reveals that $S(E)$ was correctly
reconstructed throughout the explored energy range [-1.93; 1.93]
(exploiting the ferro/antiferro symmetry) with an uncertainty that is
approximately constant (its average being 0.17) and in agreement with
the one expected {\em a priori} from eq. \ref{eqn:error}.  The
corresponding average relative error on $S(E)$ was 0.05\%, similar to
that obtained with a comparable number of sweeps in the recent and
powerful approaches described in refs. \cite{muca,wl01}.

Analogous runs were repeated for the 50x50 system using three
temperatures, $T=2.6, 4$ and $12$ and the same parameters as before.
By using 2.2\,$10^6$ Monte Carlo sweeps $S(E)$ was reconstructed over
the energy range [-1.8,1.8] again with an average error of 0.24, again
in agreement with the one expected {\em a priori}. This confirms that
the proposed strategy allows good control over the final accuracy on
$S$.


We wish to point out that, in the high-temperature limit, our approach
has strong analogies with the Wang and Landau algorithm, in which
$D(E)$ is also modified in a history-dependent fashion\cite{wl01}
(their ``pointwise'' modification of the density of states can be
viewed as a limiting case of our Gaussians). As in their case, with
one metadynamics run at a single temperature we could explore the
whole energy range.  This, however, may be inefficient, especially in
a realistic model since it could require an impractically large number
of Gaussians.  Within our approach, it is not necessary to renounce to
the Boltzmann bias, and it is possible to focalize the effort for
exploring the region of phase space of relevance for the temperature
of interest.  With this respect, our metodology can be viewed as a
finite temperature extension of the Wang and Landau algorithm.

Although in its present formulation the proposed method allows an
accurate and efficient recovery of a system entropy, it is certainly
susceptible to further generalizations and improvements. In
particular, in order to improve the resolution, the height and/or
width of the Gaussians may be changed as the metadynamics progresses,
in analogy with the method of ref. \cite{wl01}.  The application
of the method to first order phase transitions is conceptually
straightforward although, in practice, the elimination of hysteretic
effects in the metadynamics may prove computationally expensive.
However, these effects could be eliminated by exploiting the ability
of metadynamics to sample multidimensional free energy
surfaces\cite{hills}. Supplementing $E$ with auxiliary order
parameters suitable for characterizing the transition should
facilitate the overcoming of the free energy barriers associated with
the nucleation of the new phase and thus eliminate/reduce the
hysteresis.  The progress made here constitutes a substantial
improvement to the accuracy of the metadynamics approach and
illustrates its relation to other very powerful methods like multiple
histogram reweighting\cite{multihisto} and Wang and
Landau algorithm\cite{wl01}. Given the potential range of
applications of metadynamics we expect that our work will have an
impact far broader than the present demonstrative calculation on the
Ising model.


\begin{thebibliography}{99}
\bibitem{note_binomial} From the binomial statistics it follows that
the uncertainty on the number of hits collected in e.g. $E_1$ is
$\sigma_1=\sqrt{n\,p_1(1-p_1)}$. Therefore, as a measure of the
uncertainty of the thermodynamic force we have taken $\sigma ={T
\over (E_2 -E_1)} \max[ \ln{p_1 - \sigma_1 \over p_2 + \sigma_2 },
\ln{p_1 + \sigma_1 \over p_2 -\sigma_2 } ].$
\end{thebibliography}
\end{document}